\documentclass{iau} 
\usepackage{graphicx}

\title[IAUS340. CMEs as a new indicator of the active Sun ] 
{Coronal mass ejections as a new indicator of the active Sun}

\author[Nat Gopalswamy]   
{Nat Gopalswamy$^1$}

\affiliation{$^1$ Solar Physics Laboratory, NASA Goddard Space Flight Center, Greenbelt, MD 20771, USA
 \\ email: {\tt nat.gopalswamy@nasa.gov} \\[\affilskip]}

\pubyear{2018}
\volume{340}  
\jname{Long – Term Datasets for the Understanding of Solar and Stellar Magnetic Cycle}
\editors{D. Banerjee, J. Jiang, K. Kusano \& S. Solanki, eds.}
\begin{document}

\maketitle

\begin{abstract}
Coronal mass ejections (CMEs) have become one of the key indicators of solar activity, especially in terms of the consequences of the transient events in the heliosphere. Although CMEs are closely related to the sunspot number (SSN), they are also related to other closed magnetic regions on the Sun such as quiescent filament regions. This makes CMEs a better indicator of solar activity. While sunspots mainly represent the toroidal component of solar magnetism, quiescent filaments (and hence CMEs associated with them) connect the toroidal and poloidal components via the rush-to-the-pole (RTTP) phenomenon. Taking the end of RTTP in each hemisphere as an indicator of solar polarity reversal, it is shown that the north-south reversal asymmetry has a quasi-periodicity of 3-5 solar cycles. Focusing on the geospace consequences of CMEs, it is shown that the maximum CME speeds averaged over Carrington rotation period show good correlation with geomagnetic activity indices such as $Dst$ and $aa$.

\end{abstract}

\firstsection 
\section{Introduction}
The Sunspot number (SSN) is a widely used index of solar activity, representing the toroidal component of solar magnetism. The 11-year sunspot cycle is one of the key signatures of the solar dynamo. Sunspots have been systematically observed since the invention of the telescope in the 1600s. Coronal mass ejections (CMEs), on the other hand, were discovered only in 1971 with systematic observations beginning a decade later. CMEs have a wider implications for the solar dynamo because they reflect transient activity from sunspot and non-spot (quiescent filament) magnetic regions. The discovery that eruptive prominences/filaments are substructures of CMEs has helped extend CME studies back to the mid-1800s.  Since energetic CMEs have a wide variety of interplanetary consequences, we have the opportunity to infer such CMEs from historic geomagnetic storms, solar energetic particle (SEP) events, and in natural archives such as tree rings and ice cores (e.g., Usoskin 2017).

The Solar and Heliospheric Observatory (SOHO) mission has produced the longest and most uniform data base on CMEs extending over two solar cycles, providing the opportunity to investigate the relationship between CME occurrence rate and SSN over two solar cycles. Sunspot regions produce the most energetic CMEs because of the higher magnetic energy available.  CMEs from non-spot regions provide interesting information on the quiescent filaments, which occur at mid-latitudes roughly parallel to the sunspot butterfly diagram. During the maximum phase, they occur at high latitudes (polar crown filaments, see e.g., Ananthakrishnan, 1952). Cessation of high-latitude prominences occurs towards the end of the solar maximum phase and signals the sign reversal at solar poles (Babcock 1959; Waldmeier 1960; Hyder 1965). Making use of the historical prominence data available since 1860, we investigate the north-south asymmetry in the polarity reversal, which is unusual in cycle 24 (Gopalswamy et al. 2016 and references therein). There have been attempts to use CME properties to characterize the solar activity:  relationship of CME occurrence rate to SSN (Hildner et al. 1976), the fraction of halo CMEs as an indicator of the energy of a CME population (Gopalswamy 2006), and the relationship of CME maximum speed with geomagnetic variability measured by an index such as $Ap$ and $Dst$ (Kilcik et al. 2011).

\begin{figure}[t]
\begin{center}
 \includegraphics[width=4.in]{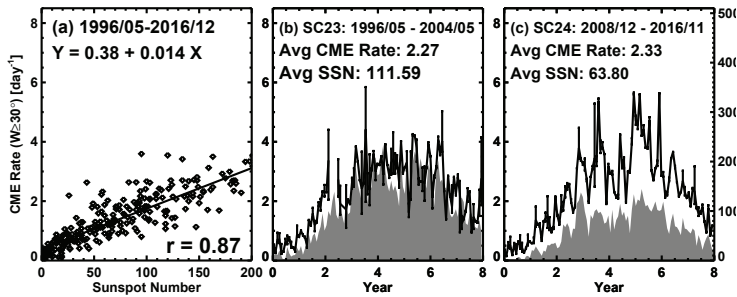} 
\caption{(a) Correlation between SSN and CME occurrence rate per day. (b) SSN (gray) and CME rate as a function of time in cycle 23, and (c) SSN and CME rate over the corresponding epoch in cycle 24. CME data are from the CME catalog: https://cdaw.gsfc.nasa.gov (Yashiro et al. 2004; Gopalswamy et al. 2009).}
   \label{fig1}
\end{center}
\end{figure}

\section{Sunspot Number and CME Occurrence Rate}
Hildner et al. (1976) first noted that the CME rate is correlated with the sunspot number using Skylab CMEs. This was confirmed with larger number of CMEs observed by coronagraphs on board P78-1 (Webb 1991; Webb and Howard 1994) and SOHO (Gopalswamy et al. 2003a) missions. One of the important findings in the SOHO era is the reduced SSN-CME rate correlation during the solar maximum phase (Gopalswamy et al. 2010) owing to the large number of CMEs from quiescent filament regions outside of the active region belt. Figure 1 shows that the high correlation between SSN and CME rate for solar cycles 23 and 24 (correlation coefficient $r$ = 0.87) with a regression line, $Y$ = 0.38 + 0.014$X$ ($X$ is SSN and $Y$ is the CME daily rate). Figure 1 also compares the CME rate and SSN between cycles 23 and 24. Clearly there is inter-cycle variation of the correlation. In cycle 24, the CME rate normalized to SSN is a factor two higher: 0.04 vs.  0.02 in cycle 23. Accordingly, the regression line for cycle 24, viz., $Y$ = 0.22 + 0.017$X$ has a higher slope than the one for cycle 23: $Y$ = 0.41 + 0.013$X$.  When the correlation analysis is performed for the individual phases of solar cycle 23, correlation coefficients show a significant difference: 0.88 (rise phase), 0.69 (maximum phase) and 0.82 (decay phase), confirming the earlier result (Gopalswamy et al. 2010). Currently the solar activity is in the decay phase of cycle 24, so we repeated the correlation analysis. We find that r = 0.85 correlation for the whole cycle, while for the rise, maximum and decay phases, it is 0.80, 0.67, and 0.72, respectively. The cycle-24 correlation is also the lowest during the maximum phase. We see that the CME rate provides information on both sunspot and non-spot magnetic regions on the Sun. 
 
  \begin{figure}[]
\begin{center}
 \includegraphics[width=4. in]{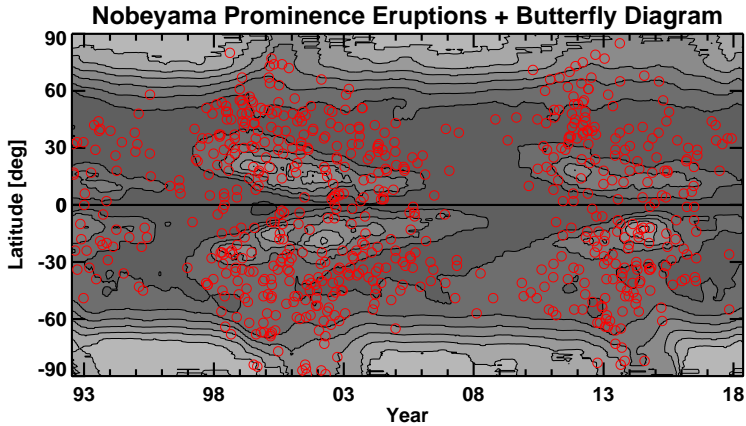} 
 \caption{Locations of prominence eruptions detected automatically from 17 GHz microwave images of the Sun obtained by the Nobeyama Radioheliograph (red circles) superposed on the microwave butterﬂy diagram (contours and gray scale). A 13-rotation smoothing has been applied to the microwave brightness temperature along the time axis to eliminate the periodic variation due to solar $B0$-angle variation. The contour levels are at [1.0, 1.03, 1.06, 1.09, 1.13, 1.16, and 1.19]$\times$10$^4$ K. The high-latitude patches correspond to times of enhanced magnetic ﬁeld strength during solar minima. The low-latitude patches correspond to active regions that become prominent during solar maxima (updated from Gopalswamy et al. 2012).}
   \label{fig2}
\end{center}
\end{figure}

\section{Rush-to-the-Pole Phenomenon and Solar Polarity Reversal}
The non-spot sources of CMEs are the quiescent filament regions that do not contain sunspots (see Fig. 2). Quiescent filaments/prominences erupt and the eruptive prominences become the inner core of CMEs appearing as bright features in white-light coronagraph images (Gopalswamy et al. 2003b). The locations of prominence eruptions (PEs) have been used as proxies to prominence locations (Gopalswamy et al. 2003c). Figure 2 shows two key features. 1. PE locations occupy a band of mid-latitudes above the active region belt delineated by the low-latitude (LL) microwave brightness contours. They even have a butterfly appearance, except that they have a latitude offset from the sunspot regions. 2. They appear at latitudes greater than 60 degrees only during the maximum phase and reach the highest latitudes typically around the time of peak SSN.  The appearance of PEs at higher and higher latitudes is known as “rush to the poles” (RTTP) phenomenon known since the late 1800s (Ananthakrishnan 1952 and references therein).  The end of RTTP and the cessation high-latitude (HL) eruptive activity mark the time of polarity reversal, which happens at different times in the two hemispheres. This was already noted when the polarity reversal was discovered in solar cycle 19 (Babcock 1959; Hyder 1965). The HL microwave brightness contours in Fig. 2 represent the polar magnetic field strength (Gopalswamy et al. 2012), which starts increasing after the end of RTTP and peaks during solar minima. The polar field strength is indicative of the poloidal magnetic field component of the solar dynamo.

\begin{figure}[]
\begin{center}
 \includegraphics[width=4 in]{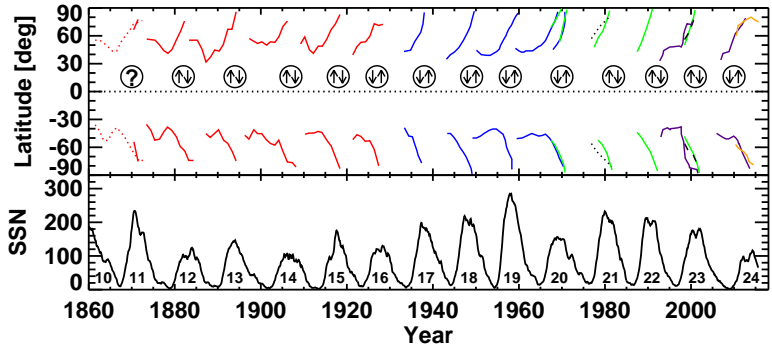} 
 \caption{RTTP traces during solar cycles 10-24 from various sources compiled by Stix (1974) (red and blue lines) with additions from Lorenc et al. (2003) (green lines), Fujimori (1984) (dotted black lines), and Pojoga and Huang (2003) (dashed black lines). The arrows inside circles indicate at which order the RTTP ended in the northern (arrow up) and southern (arrow down) hemispheres. Question mark indicates no reversal asymmetry (RTTP ends are simultaneous in the north and south).}
   \label{fig3}
\end{center}
\end{figure}

The temporal coincidence between RTTP end and the polarity reversal was further explored by Stix (1974) by compiling RTTP data all the way back to solar cycle 10. We recently used his compilation by combining it with recent data to study the north-south asymmetry in the polarity reversal (see Fig. 3, Gopalswamy et al. 2018).  It was found that the reversal asymmetry has a quasiperiodicity of 3-5 solar cycles. The reversal switched between hemipsheres during the maxima of cycles 16, 21, and 24. The reversal asymmetry has been attributed to the hemispheric asymmetry of sunspot activity (Svalgaard and Kamide 2013).  During cycle 24, the sunspot activity peaked first in the northern hemisphere, and then in the southern hemisphere. However, the reversal happened first in the south and then in the north. This is contradictory to the suggestion by Svalgaard and Kamide (2013). What happened was that the reversal in the north was delayed by the series of alternating positive and negative flux surges from the northern hemispheric active region belt that cancelled or added to the incumbent polar flux before completely reversing it in 2015. In the south, there was one big surge of negative polarity flux that was large enough to cancel the incumbent flux and quickly reversed the south pole in the year 2014 (see Gopalswamy et al. 2016 and references therein).  Thus, the sunspot asymmetry did not translate into reversal asymmetry. It is true that the reversal asymmetry is connected to the solar activity, but depends on the tilt of the emerging sunspot groups. Sunspot groups violating Joy’s law or Hale’s law contribute to the delay in the reversal and hence affect the phase and amplitude of the next cycle (Nagy et al. 2017 and references therein).

\section{CMEs and Indices of Geomagnetic Activity}
Geomagnetic indices characterize the variability of Earth's magnetic field using a single number (e.g., $Dst$, $K$, $Ap$, $aa$, and others).  Kilcik et al. (2011) considered the correlation of monthly averaged maximal CME speeds in cycle 23 with $Dst$ and $Ap$ indices and found a significant correlation. Both CME speed and SSN were correlated with $Dst$ and $Ap$, but the CME correlations were higher. This is because CMEs are the ones physically arriving at Earth causing the geomagnetic activity. Corotating interaction regions (CIRs) can also be sources of geomagnetic activity. CIR storms are more common in the declining phase of the cycle, when the fast wind from low-latitude coronal holes compresses the preceding slow wind to form CIRs. CIRs have high magnetic field strength that can lead to intense storms if the field has an out-of-the-ecliptic component that is south pointing. On the other hand, CMEs occur abundantly in the maximum phase, so one expects intense CME-related storms to occur in the maximum phase. The magnitude of the field strengths in interplanetary CMEs and CIRs are similar, but the CME speeds can be much larger (see e.g., Gopalswamy 2008). Therefore, the severest of storms are CME-related. CIR-related storms are generally milder, but more frequent. 
Considering only intense storms, Gopalswamy (2010) obtained an empirical relation, $Dst$ = $-$0.01$V$$\mathopen|B_z\mathclose|$$-$32 nT, where $B_z$ is the magnitude of south-pointing interplanetary magnetic field and $V$ is the speed of the interplanetary magnetic structure. Both $V$ and $B_z$ are ultimately related to the properties of the source region on the Sun. In particular, the CME speed is correlated with the total reconnected (RC) flux during an eruption (Gopalswamy et al. 2017; Qiu et al. 2007). For force-free flux ropes, the magnetic field strength in the flux rope is directly related to the flux-rope radius, length, and the RC flux. Thus, faster CMEs have a higher magnetic content, making the CME speed as a key parameter.

\begin{figure}[]
\begin{center}
 \includegraphics[width=4. in]{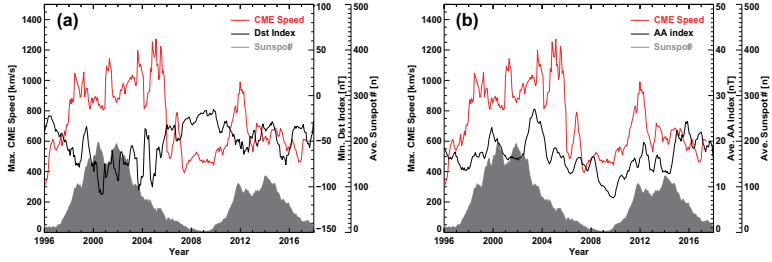} 
 \caption{(a) Maximum CME speed, minimum $Dst$, and SSN. (b) Same as in (a), except the $Dst$ curve is replaced by the $aa$ curve. Plotted are daily values averaged over Carrington rotation periods and smoothed over seven rotations.}
   \label{fig4}
\end{center}
\end{figure}
Figure 4 compares the maximum CME speed with $Dst$ index and $aa$ index, with SSN given for reference. During periods of high CME speed we see generally a more negative $Dst$ index (high negative $Dst$ means severe geomagnetic storm). We do not expect perfect correlation because a high-speed CME without a southward magnetic field component cannot produce geomagnetic activity. The $aa$ index is a positive quantity and is generally correlated with the CME speed index. There are two lowest values of $aa$ index, one in 1997 (cycle 23) and the other in 2009 (cycle 24). The lowest values are delayed by about one year from the sunspot minimum and have been used for the prediction of solar-cycle strength using the precursor method (see e.g., Wang and Sheeley and references therein). 

When there is no geomagnetic activity, the $Dst$ index is close to zero and the $aa$ index attains a minimum value. Many authors have used the $aa$ index to infer the polar field strength of the Sun and hence the amplitude of the next solar cycle under the precursor method (e.g., Wang and Sheeley 2009; Petrovay 2010 and references therein). In particular, the minimum value of the $aa$ index ($aa_{min}$) that occurs at solar-cycle minimum has been found to be a good indicator of the maximum SSN ($R_{max}$) during the upcoming solar maximum. Wang and Sheeley (2009) found a correlation coefficient of 0.93 with a regression equation, $R_{max}$ = 10$aa_{min}$$-$37.5 for cycles 12-23. The high correlation can be traced to the fact that at the activity minimum the interplanetary magnetic field at Earth originates from the solar polar region and hence $aa_{min}$ is a direct measure of the polar field strength. Thus knowing $aa_{min}$ one can predict $R_{max}$ likely to be attained half a solar cycle later. While $aa_{min}$ can predict $R_{max}$ for the whole Sun, the microwave brightness temperature at the solar poles (see Fig. 2) can predict the strength of individual hemispheres (Gopalswamy et al. 2018). Furthermore, one has to wait past the sunspot minimum to get $aa_{min}$. On the other hand, polar microwave brightness temperature has a broad plateau around the minimum, allowing for an earlier prediction of the upcoming solar cycle strength. 

\section{Discussion and Summary}
We considered the CR-averaged occurrence rate and peak speeds of CMEs as new indicators of solar activity and compared them with the sunspot number. CMEs provide a more comprehensive view of solar activity because they cover all closed magnetic regions (sunspot and non-spot). We also considered the RTTP phenomenon in individual hemispheres over 15 solar cycles that shows interesting behavior of the polarity reversal. Finally, we compared the CME maximum speed with $Dst$ and $aa$ index. The main conclusions of the investigations are the following.
1. The CME occurrence rate is closely correlated with SSN. The correlation is weaker during the maximum phase because of the non-spot sources of CMEs (quiescent filament regions) that are abundant during the maximum phase.\\ 
2. The locations of prominence eruptions serve as proxy to the RTTP phenomenon. The end of RTTP is a good indicator of sign reversal at solar poles and hence good indicator of the north-south asymmetry in the polarity reversal.\\ 
3. Examination of the RTTP phenomenon over the 15 solar cycles since 1860 indicates that the reversal asymmetry has a quasi periodicity (3-5 cycles). One of the possible reasons is the prolonged presence of high-latitude prominences due to a series of flux surges of opposite polarity that delay the reversal.\\
4. The maximum CME speed is a good indicator of the level of geomagnetic activity and hence can be used as a solar activity index. The speed and the magnetic content of CMEs are determined by the reconnected flux in the eruption region.\\
5. Unlike the $Dst$ index, the $aa$ index is highly useful in solar cycle prediction. The polar microwave brightness has a similar prediction value, except that it can predict the strength of the cycle in individual hemispheres.

{\underline{\it Acknowledgement}}.
I thank S. Akiyama, S. Yashiro, and P. M{\"a}kel{\"a} for help with figures, and B. Prabhu Ramkumar, R. Selvendran, and C. Ebenezer for help in examining Kodaikanal Solar Observatory prominence drawings. This work benefited from the open data policy of NASA/ESA (SOHO CME data), SILSO (sunspot data), ICCON (Nobeyama Radioheliograph data),  NGDC ($aa$ index), Kyoto University ($Dst$ index) Work supported by NASA Heliophysics GI and LWS programs.


\begin{thebibliography}{}

\bibitem[Ananthakrishnan(1952)]{1952Natur.170..156A} Ananthakrishnan, R.\ 1952, \textit{Nature}, 170, 156 

\bibitem[Babcock(1959)]{1959ApJ...130..364B} Babcock, H.~D.\ 1959, \textit{ApJ}, 130, 364 

\bibitem[Fujimori(1984)]{1984PASJ...36..189F} Fujimori, K.-I.\ 1984, \textit{PASJ}, 36, 189 

\bibitem[Gopalswamy(2006)]{2006GMS...165..207G} Gopalswamy, N.\ 2006, \textit{AGU Geophysical Monograph Series}, 165, 207 

\bibitem[Gopalswamy(2008)]{2008JASTP..70.2078G} Gopalswamy, N.\ 2008, \textit{JASTP}, 70, 2078 

\bibitem[Gopalswamy(2010)]{2010IAUS..264..326G} Gopalswamy, N.\ 2010, Solar \& Stellar Variability: Impact on Earth and Planets, \textit{IAUS}, 264, 326 

\bibitem[Gopalswamy et al.(2003a)]{2003ESASP.535..403G} Gopalswamy, N., Lara, A., Yashiro, S., Nunes, S., \& Howard, R.~A.\ 2003a, Solar Variability as an Input to the Earth's Environment, \textit{ESA-SP}, 535, 403 

\bibitem[Gopalswamy et al.(2003b)]{2003ApJ...586..562G} Gopalswamy, N., Shimojo, M., Lu, W., et al.\ 2003b, \textit{ApJ}, 586, 562 

\bibitem[Gopalswamy et al.(2003c)]{2003ApJ...598L..63G} Gopalswamy, N., Lara, A., Yashiro, S., \& Howard, R.~A.\ 2003c, \textit{ApJ}, 598, L63

\bibitem[Gopalswamy et al.(2009)]{2009EM&P..104..295G} Gopalswamy, N., Yashiro, S., Michalek, G., et al.\ 2009, \textit{Earth Moon and Planets}, 104, 295 

\bibitem[Gopalswamy et al.(2010)]{2010ASSP...19..289G} Gopalswamy, N., Akiyama, S., Yashiro, S., \& M{\"a}kel{\"a}, P.\ 2010, \textit{Astrophysics and Space Science Proceedings}, 19, 289 

\bibitem[Gopalswamy et al.(2012)]{2012ApJ...750L..42G} Gopalswamy, N., Yashiro, S., M{\"a}kel{\"a}, P., et al.\ 2012, \textit{ApJ}, 750, L42
 
\bibitem[Gopalswamy et al.(2016)]{2016ApJ...823L..15G} Gopalswamy, N., Yashiro, S., \& Akiyama, S.\ 2016, \textit{ApJ}, 823, L15 

\bibitem[Gopalswamy et al.(2017)]{2017arXiv170508912G} Gopalswamy, N., Akiyama, S., Yashiro, S., \& Xie, H.\ 2017, \textit{JASTP},  arXiv:1705.08912

\bibitem[Gopalswamy et al.(2018)]{2018arXiv180402544G} Gopalswamy, N., M{\"a}kel{\"a}, P., Yashiro, S., \& Akiyama, S.\ 2018, arXiv:1804.02544 

\bibitem[Hildner et al.(1976)]{1976SoPh...48..127H} Hildner, E., Gosling, J.~T., MacQueen, R.~M., et al.\ 1976, \textit{Solar Phys.}, 48, 127 

\bibitem[Hyder(1965)]{1965ApJ...141..272H} Hyder, C.~L.\ 1965, \textit{ApJ}, 141, 272 

\bibitem[Kilcik et al.(2011)]{2011ApJ...731...30K} Kilcik, A., Yurchyshyn, V.~B., Abramenko, V., et al.\ 2011, \textit{ApJ}, 731, 30 

\bibitem[Lorenc et al.(2003)]{2003ESASP.535..129L} Lorenc, M., Pastorek, L., \& Rybansk{\'y}, M.\ 2003, Solar Variability as an Input to the Earth's Environment, \textit{ESA-SP}, 535, 129 

\bibitem[Nagy et al.(2017)]{2017SoPh..292..167N} Nagy, M., Lemerle, A., Labonville, F., Petrovay, K., \& Charbonneau, P.\ 2017, \textit{Solar Phys.}, 292, 167 

\bibitem[Petrovay(2010)]{2010LRSP....7....6P} Petrovay, K.\ 2010, \textit{Living Reviews in Solar Physics}, 7, 6 

\bibitem[Pojoga \& Huang(2003)]{2003AdSpR..32.2641P} Pojoga, S., \& Huang, T.~S.\ 2003, \textit{Adv. Space Res.}, 32, 2641 

\bibitem[Qiu et al.(2007)]{2007ApJ...659..758Q} Qiu, J., Hu, Q., Howard, T.~A., \& Yurchyshyn, V.~B.\ 2007,\textit{ApJ}, 659, 758 

\bibitem[Stix(1974)]{1974A&A....37..121S} Stix, M.\ 1974, \textit{A\&A}, 37, 121 

\bibitem[Svalgaard \& Kamide(2013)]{2013ApJ...763...23S} Svalgaard, L., \& Kamide, Y.\ 2013, \textit{ApJ}, 763, 23 

\bibitem[Usoskin(2017)]{2017LRSP...14....3U} Usoskin, I.~G.\ 2017, \textit{Living Reviews in Solar Physics}, 14, 3 

\bibitem[Waldmeier(1960)]{1960ZA.....49..176W} Waldmeier, M.\ 1960, \textit{Z. Astrophys.}, 49, 176

\bibitem[Wang \& Sheeley(2009)]{2009ApJ...694L..11W} Wang, Y.-M., \& Sheeley, N.~R.\ 2009, \textit{ApJ}, 694, L11 

\bibitem[Webb(1991)]{1991AdSpR..11...37W} Webb, D.~F.\ 1991, \textit{Adv. Space Res.}, 11, 37 

\bibitem[Webb \& Howard(1994)]{1994JGR....99.4201W} Webb, D.~F., \& Howard, R.~A.\ 1994, \textit{JGR}, 99, 4201 

\bibitem[Yashiro et al.(2004)]{2004JGRA..109.7105Y} Yashiro, S., Gopalswamy, N., Michalek, G., et al.\ 2004, \textit{JGR}, 109, A07105 


\end{thebibliography}
\end{document}